# Computing, Information Systems & Development Informatics Journal

## Volume 3. No. 3. July, 2012

# A Secure Intelligent Decision Support System for Prescribing Medication

**Omotosho A.**
Dept. of Computer Science and Technology
Bells University of Technology
Ota, Ogun State. Nigeria
bayosite2000@yahoo.com

**Emuoyibofarhe .O.J**
Dept. of Computer Science and Engineering
Ladoke Akintola University of Technology
Ogbomoso, Oyo State. Nigeria.
eojustice@gmail.com

Reference Format:   Omotosho .A. & Emuoyibofarhe .O.J (2012). A Secure Intelligent Decision Support System for Prescribing Medication. Computing, Information Systems & Development Informatics Journal. Vol 3, No.3. pp 9-18. Available online at www.cisdijournal.net

# A Secure Intelligent Decision Support System for Prescribing Medication

Omotosho .A. & Emuoyibofarhe .O.J

**ABSTRACT**

The process of electronic approach to writing and sending medical prescription promises to improve patient safety, health outcomes, maintaining patients' privacy, promoting clinician acceptance and prescription security when compared with the customary paper method. Traditionally, medical prescriptions are typically handwritten or printed on paper and hand-delivered to pharmacists. Paper-based medical prescriptions are generating major concerns as the incidences of prescription errors have been increasing and causing minor to serious problems to patients, including deaths. In this paper, intelligent e-prescription model that comprises a knowledge base of drug details and an inference engine that can help in decision making when writing a prescription was developed. The research implements the e-prescription model with multifactor authentication techniques which comprises password and biometric technology. Microsoft Visual Studio 2008, using C# programming language, and Microsoft SQL Server 2005 database were employed in developing the system's front end and back end respectively. This work implements a knowledge base to the e-prescription system which has added intelligence for validating doctor's prescription and also added security feature to the e-prescription system..

## 1.0 INTRODUCTION

Medical prescriptions are typically handwritten and sometimes through the use of a pre-formatted printed paper. Studies have shown that fatal health problems could arise due to Adverse Drug Effect (ADE) resulting from erroneous prescription, illegibly written prescriptions, errors in dosage and unanticipated drug interactions, communication errors committed during ordering, dispensing and administering of drugs, and dosing mistakes such as incorrect dose of drug and incorrect frequency of drug intake, and lack of reliable health information [7]. The purpose of Information and Communication Technologies (ICT) for health (also known as e-Health) is to improve significantly the quality, access and efficacy of healthcare for all individuals. ICT for health describes the application of information and communication technologies in a way that affects the whole range of activities that affect and supports the health sector. With e-Health the quality of health care can be improved while at the same time controlling escalating costs.

E-prescribing enables health care providers to electronically generate and submit prescriptions directly to a pharmacist. An e-prescribing system also allows providers to evaluate a patient's medication history, allergies, possible drug interactions, and drug coverage information. This can ensure that informed choices are made for patients. Pharmacies can also communicate with physicians through e-prescribing systems to clarify prescription orders and process renewal requests. E-prescribing is conducted by physicians through technology that is available for use within their practice.

These include stand-alone e-prescribing systems, which work on a desktop computer, laptop, or Personal Digital Assistant (PDA), and e-prescribing applications that are part of Electronic Health Record (EHR) systems. Implementing a stand-alone system automates the prescribing process and allows physicians to store and manage prescription information, and some physicians have used this as a first step toward broader IT use in their practice [4]. E-Health and most especially, e-Prescribing and medication management play important roles in improving health services provision to patients. E-prescribing enables not only faster and more reliable services but most importantly, more security and less medication errors. Overall, it unarguably contributes to patient safety. E-Prescribing contributes to patient safety because it eradicates problems associated with handwriting illegibility and, when combined with decision-support tools, automatically alerts physicians, when prescribing, and pharmacists, when dispensing, to possible interactions, and other potential problems.

In [9] , it was noted that Community Pharmacists have been developing computer assisted practices, including specific applications that enable and facilitate medication management and the provision of information as well as reporting and learning systems that enhance the identification and prevention of errors. This course of action works as a facilitating factor to implementing electronic prescribing, with synergies resulting from such a combination leading to improved patient safety and, thus, care. Medication management is especially important in the treatment of chronic conditions such as asthma, hypertension and diabetes, where the correct use of both medicines and medical devices is vital for adherence to therapies and for deriving maximum therapeutic benefit.

## 2. RELATED WORKS

Research has shown that digitalization of the medication process causes significant reduction of administrative errors, dosing errors and over-prescriptions [1]. In the works of [10], electronic prescribing was described to have the ability not only to streamline the prescription writing process, but also to reduce the number of errors that may be incurred with hand-written prescriptions. In his paper he investigates these phenomena in the U.S.A and the approach used was that a number of hypotheses were tested using principal-components analysis (PCA) and factor analyses. Although this study appears to represent the e-prescribing process in the U.S.A., the sample size and region studied is only one slice of the general population.

Unfortunately, the adoption of e-prescribing has been difficult to attain owing to numerous barriers throughout the industry and such acceptance barriers include lack of technology trust, associated system costs, and risk of unsecure patient health and medical information. In conclusion, this article documents that increasing numbers of pharmacies today are building their IT-infrastructures to accept electronic prescriptions and it may soon be the preferred method for physicians to write prescriptions. It is with great anticipation that this technology will also enhance the prescription-writing abilities of prescribing physicians globally, giving them electronic access to patient medical records and resources that will assist them in prescribing the correct drug for the patient.

According to [8], computerized drug prescribing alerts can improve patient safety, but are often overridden because of poor specificity and alert overload. Their objective was to improve clinician acceptance of drug alerts by designing a selective set of drug alerts for the ambulatory care setting and minimizing workflow disruptions by designating only critical to high-severity alerts to be interruptive to clinician workflow. The alerts were presented to clinicians using computerized prescribing within an electronic medical record in 31 area practices. There were 18,115 drug alerts generated during our six-month study period. Of these, 12,933 (71%) were non-interruptive and 5,182 (29%) interruptive. Of the 5,182 interruptive alerts, 67% were accepted. Reasons for overrides varied for each drug alert category and provided potentially useful information for future alert improvement. These data suggest that it is possible to design computerized prescribing decision support with high rates of alert recommendation acceptance by clinicians.

Author in [11] stated that electronic prescribing and computerized drug management can improve the safety, quality and cost-effectiveness of prescribing. However, if the problems that lead to avoidable adverse events are not addressed by information technology, there is a risk of making considerable investment without the expected return of error reduction and improved patient safety. Improving the safety of prescribing is particularly important in ambulatory care, where most drugs are prescribed.

He furthered that to improve patient safety, Information Technology solutions should be developed that provide: (1) access to the list of all currently active drugs, (2) alerts for relevant prescribing problems (therapeutic duplication, excess dose, dose adjustment for weight (children, elderly) and renal impairment, drug-disease, drug-drug, drug-age and drug-allergy contraindications), (3) the capacity to electronically submit medication stop orders to the dispensing pharmacy and (4) integration of electronic prescriptions (e-rx) into pharmacy software to avoid transcription errors. To improve quality of prescribing, IT solutions should be capable of providing physicians with reminders and alerts for evidence-based preventive care and disease management based on patient-specific drug, disease, therapeutic intent and other relevant clinical information.

In [3] , it was also identified and reported that medication errors are common, and while most such errors have little potential for harm they cause substantial extra work in hospitals. A small proportion do have the potential to cause injury, and some cause preventable adverse drug events. The objective of this work is to evaluate the impact of computerized physician order entry (POE) with decision support in reducing the number of medication errors. The research employed a prospective time series analysis, with four periods. All patients admitted to three medical units were studied for seven to ten-week periods in four different years. The baseline period was before implementation of POE, and the remaining three were after. Sophistication of POE increased with each successive period. The main outcome measure of the research is medication errors, excluding missed dose errors. In conclusion, computerized POE substantially decreased the rate of non-missed-dose medication errors. A major reduction in errors was achieved with the initial version of the system, and further reductions were found with addition of decision support features.

In [5], the authors reviewed literature on e-prescription, proposed and implements a model for secured and intelligent e-Prescribing system based on gaps noted. They concluded that electronic prescribing promises to improve medical prescription writing with decision support features such as drug allergy, drug-drug interaction warnings and timely information available to physician. Likewise, it was noted that majority of piloted e-prescription projects and fully deployed systems demonstrated the wide usage of smart cards for authentication and security by the prescribing physician. The paper concluded that, though smart cards are universal, portable and safe, since they contain the holder's identity, anyone who is able to access their information can steal their identity. Hence with improved intelligence features and security measures, electronic prescribing will greatly contribute to the health industry and their outcomes.

## 3.0 MATERIALS AND METHODS

The following scientific procedures were used to achieve the central idea of this work. They are: Requirement Definition and Infrastructural.

### 3.1 Infrastructural Model and Architect: The E-Prescription System Architecture and Components

The architecture for the designed electronic prescription system was based on existing architectures and it was an improvement of the Finnish e-prescription architecture proposed in [2]. Additional components have been integrated into the system, which is intelligence and biometric security. The reason for this is to improve the prescription writing, reduce prescription error, improve prescription security and reduce prescription forgery. The software architecture of the designed e-prescription shows the structure and organization by which the system components and subsystems interact to form the actual system. The system architecture in Figure 1 shows how the e-prescription system operates.

#### 3.1.1 Central Prescription Database

When the e-Prescription model was planned, many approaches for the system architecture were identified. The decision was made to implement a model using a concentrated database where all prescriptions are stored and accessed by pharmacies, physicians and the social security authorities. The centralized e-prescription database server stores the details of all e-prescription generated and sent to all registered pharmacy. This model is based on the lowest level of technical capability of all stakeholders likely to be involved in the e-Prescription chain as it is currently modelled on face-to-face interactions between the patient and the healthcare provider and also the use of wired workstations.

The design can be extended to support Internet-based consultation and the use of wireless devices by prescribers to interact with a local e-Prescription server. One of the strong drivers for e-Prescription implementation is the provision of relevant reports on resource (medication) utilisation, provider prescription behaviours and cost effectiveness. Such information is of particular importance to fund holders and payers. The hospital-based and central repositories and the e-Prescription data warehouse will provide accurate source data for such reporting and knowledge discovery purposes. This prescription database does not reside on the hospital local network and is accessible to all registered pharmacies.

#### 3.1.2 Prescriber Authentication and Security Module
#### 3.1.2.1 Fingerprint scanner

Fingerprint biometric provides a unique identification and authentication solution compared to the use smart card found in the existing systems. Smartcards could be misplaced and misused. In order to further control access to the system, the e-prescription system uses a multi-factor authentication requirement that calls for something you know and something you are. The proposed rule enforces the prescriber authenticate to the system by providing a user id, password and fingerprint biometrics scan before the system can be accessed for electronic prescription writing and transmission.

This authentication means helps to reduce cases of prescription forgery and abuse. The fingerprint templates of the prescriber and patient are stored via a fingerprint scanner in the server as byte. The fingerprint scanner is used to register both the physician and patients. It will also be used at the pharmacy by the patient to in order to display their active prescriptions.

#### 3.1.2.2 Prescription Security

In addition to securing the prescribers login to the system, the generated electronic prescription is further secured both in the printed and electronic format by automatically appending the prescribers licence ID to the e-prescription. This ID is verified by the dispensing pharmacy to confirm the authenticity and reliability of the e-prescription source. This further helps to reduce the cases of prescription forgery at the same time delivering prescription faster.

#### 3.1.3 Intelligence Module

This consists of the knowledge base, which serves as the repository where intelligence is derived from, and the inference engine, which makes deduction based on the available knowledge. The intelligence module helps in reducing prescription error.

#### 3.1.3.1 Knowledge Base

This is the repository for patients' electronic health record patient health information generated by one or more encounters in the care delivery setting and this serves as the knowledge base to the electronic prescribing system. Included in this information are patient bio-data, problems, medications, vital signs and past medical history. The EHR automates and streamlines the clinician's workflow. The EHR has the ability to generate a complete record of a clinical patient encounter, as well as supporting other care-related activities directly or indirectly via interface— including evidence-based decision support, quality management, and outcomes reporting. The knowledge base also includes a database of drug with information for prescriber about drug. The drug database is updateable and also allowing adding of new drugs and drug information thus enlarging the drug knowledge base. Drug information includes: pharmacological class, generic description, indications, usage etc.

#### 3.1.3.2 The Inference engine

This helps in deducing a patient prescription pattern from the knowledge base. The inference engine helps with fast electronic prescription writing by fetching only the needed and appropriate data from the knowledge base. It provide the system intelligence features that assists in determining and using patient past medication pattern and histories to predict their new prescription pattern and thus reducing the possibilities of common prescription errors such as incomplete dosage.

#### 3.1.4 Application Software

This is installed on the prescribers system and allows the user of the system, i.e. the administrator and the physician to perform the basic related operations, i.e. add patient, add drugs information , generate prescription, as well as add more users to access the system (only the Administrator can add more users of the system). Prescribers gained access to the e-prescribing system by providing correct and required authentication details through the application interface.

The application software provides all the necessary tools needed to write prescription by communicating with the intelligence module.

### 3.1.5 Local Area Network Link and Firewall

The e-prescription application software can be accessed by physician terminals on the hospital intranet network so they are connected within the hospital. The intranet technology facilitates communication between physician to improve data sharing capability and over all knowledge base of the physician. The firewall helps to control access between the intranet and the internet to permit access to the intranet only to physician. The firewall permits or denies network transmissions based upon a set of rules and is frequently used to protect networks unauthorized access while permitting legitimate communications to pass.

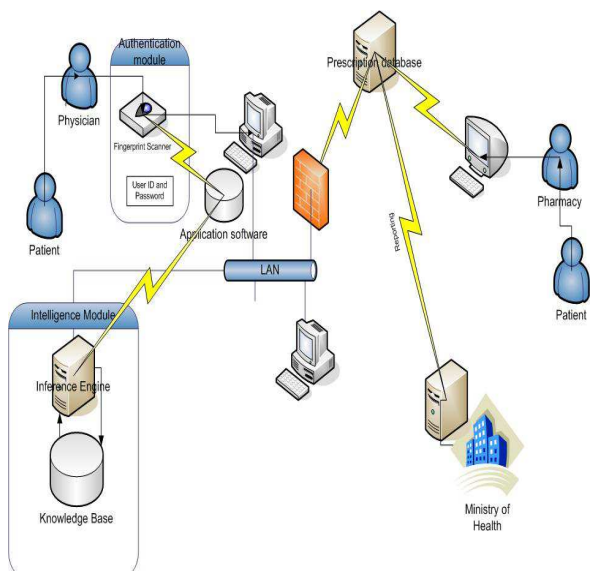

**Fig 1. System Architecture of the Proposed Electronic Prescription System**

### 3.2 Analysis of Model

The structure of the e-prescription model was identified using the Unified Modeling Language (UML) analysis model. The analysis model led to the derivation of scenario based elements. For Scenario-based elements use-case diagram and activity diagram were designed for the e-prescription system. Figure 2 shows the Use Case scenario for the proposed secured intelligent e-prescription system.

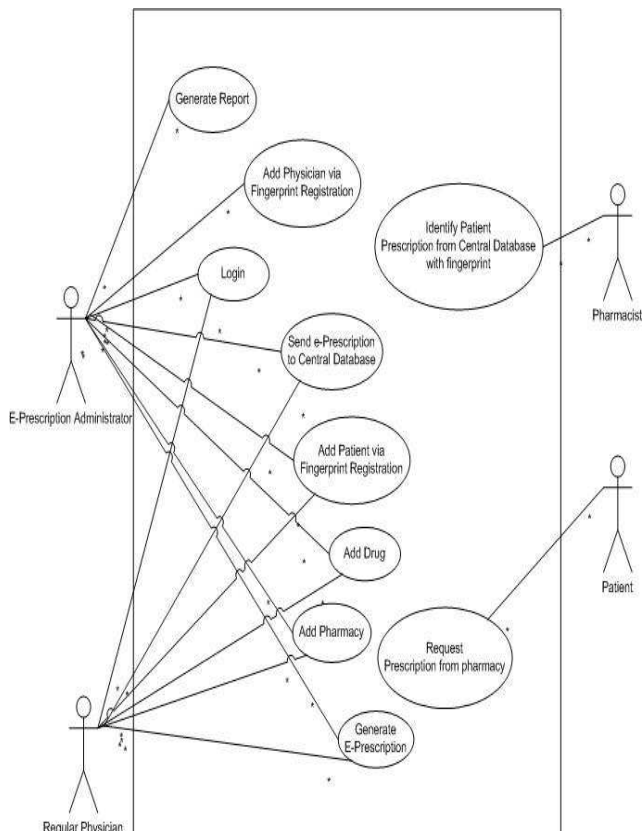

**Fig 2. Use Case scenario for the proposed secured intelligent e-prescription system[5].**

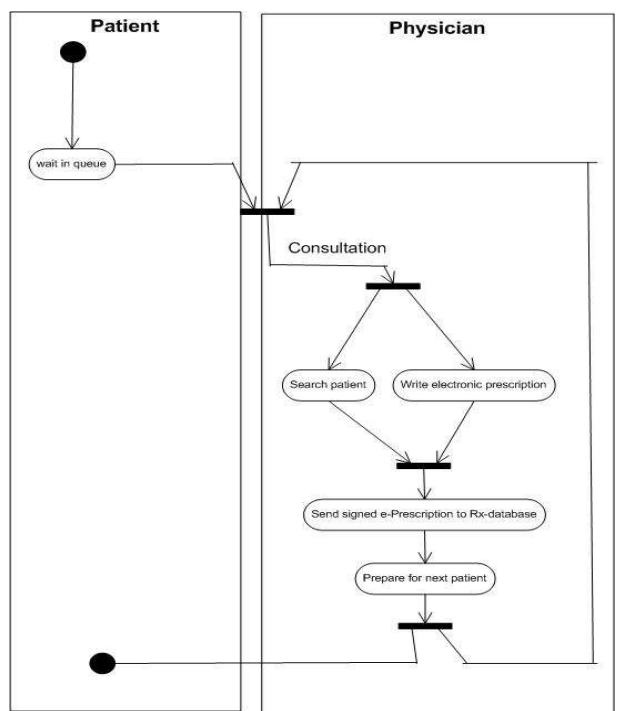

**Fig 3. Business level activity diagram of the proposed secured intelligent e-prescription system.**

In figure 2, there are four main actors: the administrative physician, the regular physician, patient and pharmacy. The registered physicians sign-in into the e-prescription system via a biometric finger print authentication and then search for patient before writing prescription. The e-prescribing system provides an intelligent feature that improves accuracy of prescription. At the pharmacy, with the use of biometric, tracking of e-prescriptions in the central prescription server are achieved. Also, the administrative physician/ regular physician can add drugs, patient, write and send prescription electronically, etc. All information here is stored in the centralized server and can be used for future reference in addition to the report accessible to the administrative physician. Figure 3 shows the electronic prescription business level activity diagram.

*3.3 Database Model of the System*
A relational modelling technology is adopted for the design of the application's database. The Relational Database Management System used in this system is Microsoft SQL Server 2005. The database model of the system is composed of the listed files below:

1. users [user_id, password, fullname, user_type, phone_no, fingerprint, prescriber_no]

2. DrugList[drug_id,name,Legal_class,Manufacturer,Pharmacological_Class, General_Description, Indications, Adult_Usage, Children_Usage, Contraindications, Precautions, Interactions, Adverse_Reactions, How_Supplied]

3. Medication [med_id, pat_id,drug_id, pat_name, med_name, num, refill, substitute, dosage, freq, route, sig, note, start_d, refill_d, renew_d, pharmacist, date]

4. Pharmacist[Pharm_id,pat_id,name, address, phone, email]

5. patient [ pat_id, use_id, fullname, phone, dob, address, drug_allergy, occupation, pharmacist]

*3.4 Knowledge Base Model of the System*
The e-prescription knowledge base system uses the knowledge of drug information and medications obtained from human experts that is, physicians for proposing medication and this was programmed into the developed system. Case Based Reasoning (CBR), the process of solving new problems based on the solutions of similar past problems, was used to build the knowledge base, the system reasons from experiences or "old cases" in an effort to suggest medications. One of the primary goals of Case Based Reasoning Systems (CSRS) is to find the most similar, or most relevant, cases for new input problems. The effectiveness of CBRS depends on the quality and quantity of cases in a case memory. Figure 4 shows the Case Retrieval and Adaptation Mechanisms. As medications are administered the e-prescription system keeps records of the different cases of successful prescription and makes them available in the medication suggestion when a patient requires similar treatment in the future.

Case comparison performed in order to produce a matching solution, for new medications prescribed for the first time, regarded as a new case, the system stores them as likely solutions when making future similar prescription as they also form part of the medication case memory. The CBR based knowledge systems uses the CBR methodology as an inference technique, an extensive past cases as a knowledge structure and solves new problems by adapting solutions of similar problems. The basic IF-THEN-RULES models used in the e-prescription CBR Inference engine design are represented in Figure 5.

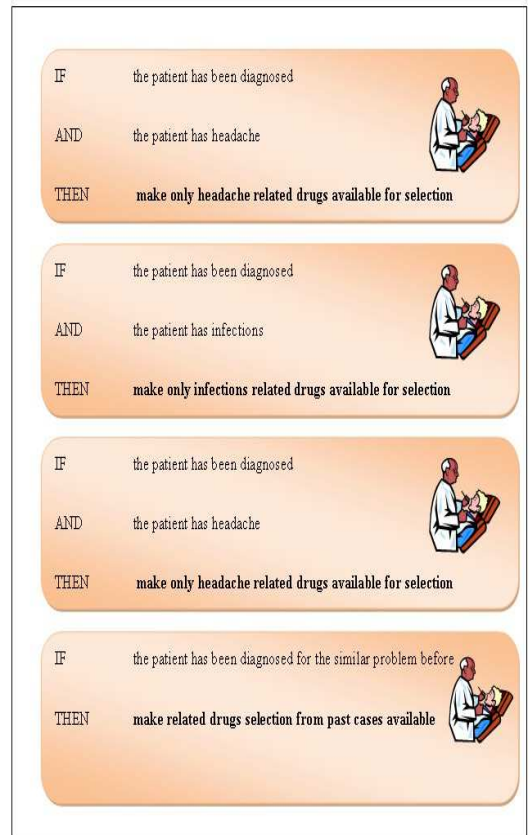

**Fig 4. The E-prescription CBR Inference Engine Design**

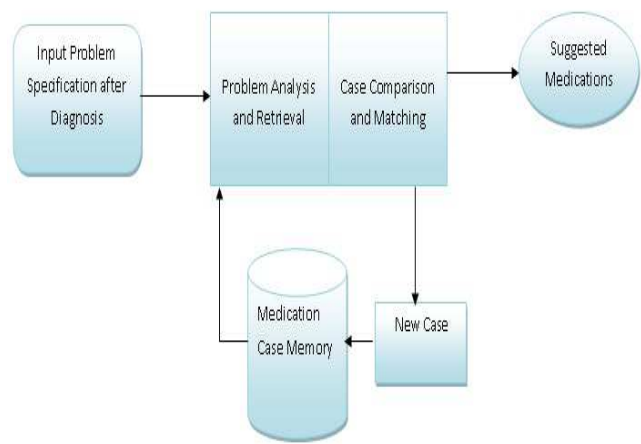

**Fig 5. Case Retrieval Mechanisms of the E-prescription Knowledge base.**

## 4.0 SYSTEM IMPLEMENTATION AND RESULTS

Microsoft Visual Studio 2008 C# programming language was employed in developing the electronic prescription application software interactive interface and Microsoft SQL Server 2005 was used to build the data store as it is one of the powerful multiuser databases that is compatible with the development language. Also, biometric fingerprint capturing hardware was programmed to work with the system. Figures below shows the impact which biometric and intelligence could have in the when used in electronic prescription.

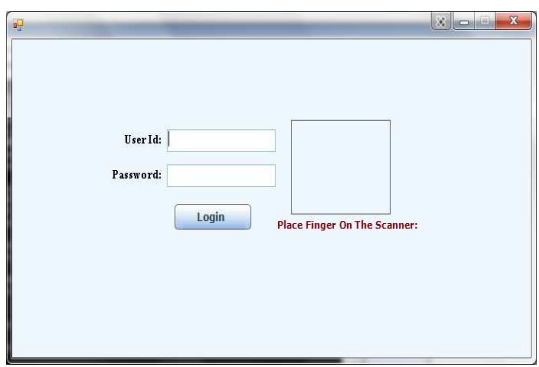

**Fig 6: Biometric Based Login page**

In Figure 6 the login interface, the system only allows authorized
physician to access it. In the login screen, a physician or administrator must type their username, password and provide their fingerprint scan through a fingerprint reader in order to access the system. The administrator who is the super user has a better privilege after his login than a regular physician. In case of a mismatch in the login credentials the system would trigger an appropriate error message.

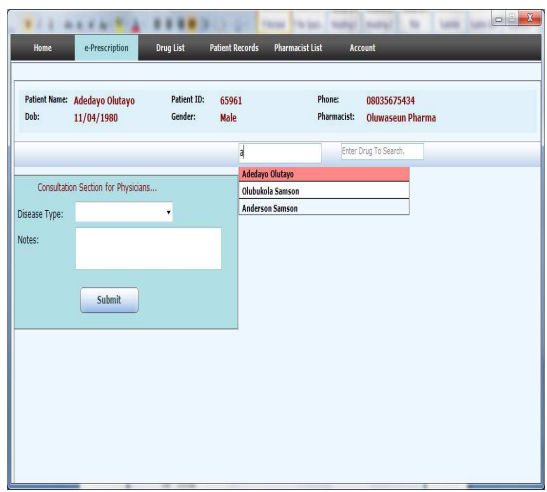

**Fig 7. Physician Consultation Page**

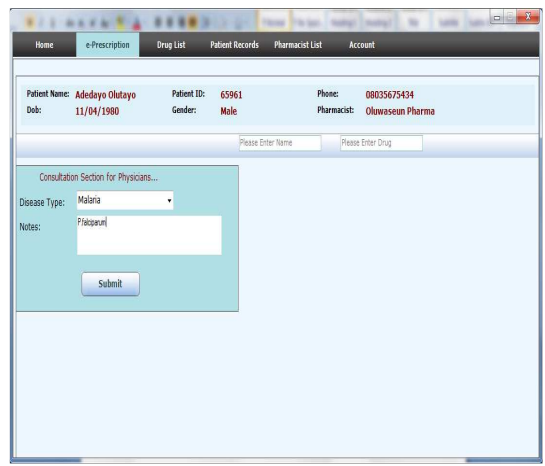

**Fig 8. Sample Consultation Note**

Figure 7 shows the physician consultation page, the prescribing physician diagnose the patient first before prescription is made, the system accepts the nature of the patients disease or infection and the physician specific description of the patient health problem after the consultation process. After the submit button is clicked, the consultation result is added to the patient medical history as this can be useful for making important future prescription decision on the patient health . Also, the patient search box allows the prescribing physician to search for registered patient prior to consultation and prescription writing.

The system provides an auto complete feature that automatically list patient names on typing the first letter of the name to be searched. The quick search of patient name helps in faster patient record retrieval. Once a patient is identified, the bio-data of the corresponding patient is immediate loaded. This feature also ease e-prescription writing as physicians do not have to waste time identifying patients within the system before prescription can be generated. Figure 8 shows a sample consultation note by physician once a patient has been identified. For example, patient Adedayo Olutayo has been diagnosed for *p.falciparum* malaria.

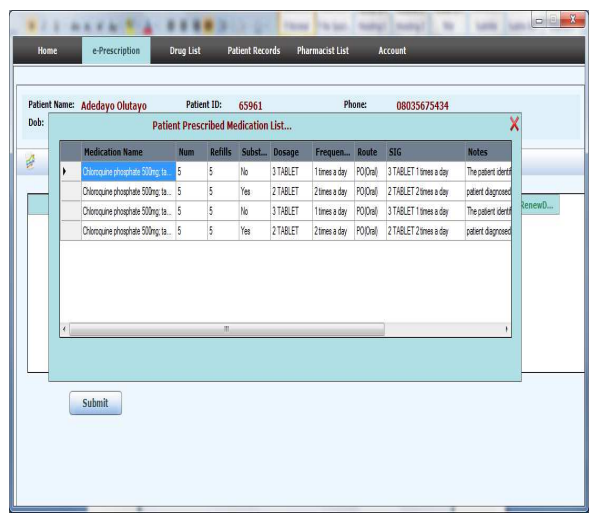

**Fig 9. Current Patient Prescription Patterns.**

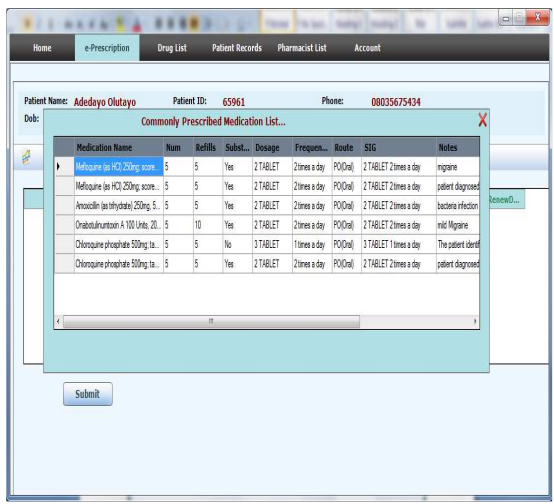

**Fig 10. Commonly Prescribed Medication List.**

Figure 9 shows list of the currently selected patient medication pattern from which new medications can be optionally provided. For example in the case of emergency, instead of searching through the entire drug list to make a new medication, e-prescription can be quickly generated based on the patient history provided by the system. The intelligent system keeps records and makes available history of successful prescriptions the current patient had taken based on the patient health record data. This feature ensures that prescriber, even if there is a need to write a new medication, does not over deviate from prescription that can treat the patient health problem thus reducing cases of prescription errors resulting in adverse drug effects.

Figure 10 displays list of recently or commonly prescribed medications. This is a list of previous prescriptions made for all patients and a new prescription can be generated from this history as well. As prescribers keep on using the system, the system automatically keeps track of the most frequently prescribed drugs along with their prescription detail. So, by selecting from this list, a complete prescription is generated without having to re-type (though it can be modified) the prescription parameters like the dosage, route of administration etc.

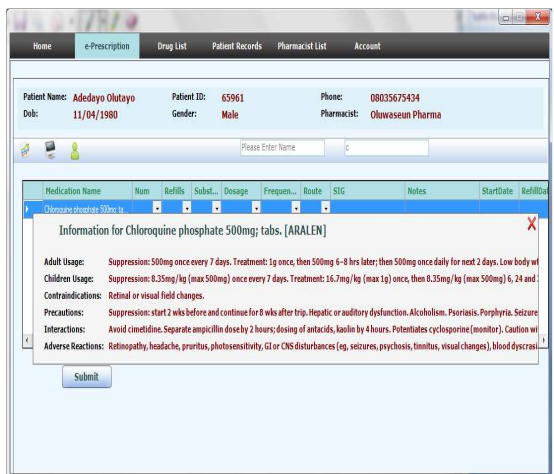

**Fig 11. e-Prescription Intelligence Drug Information Page**

Figure 11 shows the e-prescription intelligence help page. On any prescription made, by simply selecting the drug name, the system provides an intelligent feature that provides immediate information on the currently selected medication. Information provided may include drug usage, contraindications, precautions, adverse reaction etc. all of which can provide prescriber timely information on the drug they are prescribing which also affects the patients' health. Since lack of reliable health data was identified as one of the major causes of adverse drug effect, this help pop menu therefore provides up-to-date information about a drug as long as the drug information in the drug database is updateable.

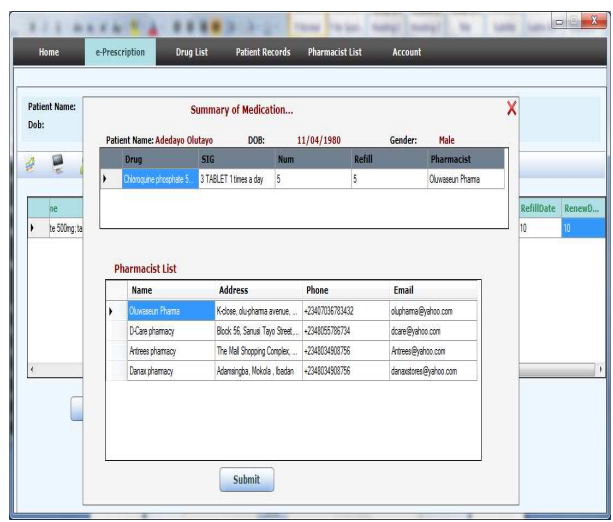

**Fig 12. Selecting pharmacy and Transmitting Prescription Electronically**

Figure 12 shows the pharmacy selection page before e-prescription transmission. On this page, physicians are provided with summary of the currently generated prescription for a final review before it is transmitted to the pharmacy. However, for a patient who did not specify pharmacy during his or her registration a pharmacy choice, which automatically becomes the default pharmacy for the patient, can be made here; otherwise if pharmacy name is specify during patient registration, the patients default pharmacy appears on the prescription. Though pharmacy choice is changeable but selection is limited to those that are registered. Once the submit button is clicked, the electronic prescription is sent to the central prescription database which is accessible to the dispensing pharmacy.

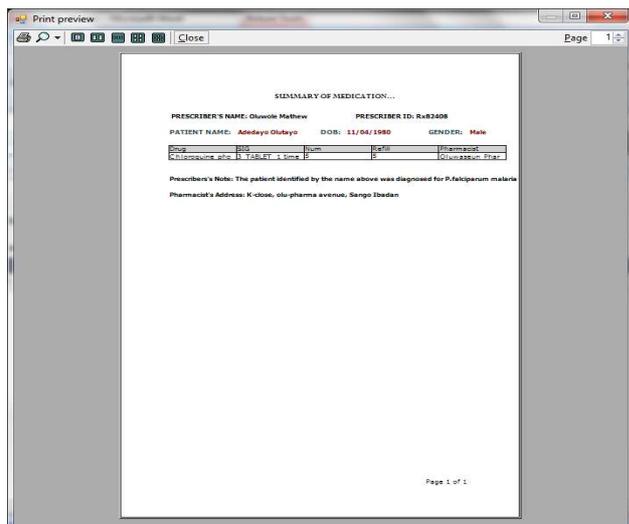

**Fig 13. Selecting pharmacy and Transmitting Prescription Electronically**

Figure 13 shows the e-Prescription final print preview as it will be appear on the printing paper. On the prescription is patient information, the prescription being made, the prescribers note, the pharmacist address and more importantly for security purpose, verification and monitoring, the prescribers name and a unique prescriber id which identifies any practicing, qualified and registered physician without which the e-prescription will not be dispensed by the receiving pharmacist. The unique prescriber id further secured the electronic prescription however; it is subject to re-verification at the pharmacy.

## 4.0 CONCLUSION

The intelligent features of the system provide an avenue to avoid common prescription errors which usually result in adverse drug effect with the provision of patient prescription patterns, timely drug information, and the prevention of incomplete dosage errors among others. Lastly in the framework of the developed system, in addition to the prevention of an unauthorized access to the e-prescribing system with the use of biometric security, the generated prescription is further secured by ratifying it with the prescribers licence number which could be verified by the pharmacist before dispensing a prescription hence, the prescription source is verifiable. This also indicates that the prescriber takes responsibility for the clinical care of the patient and the outcome. In the end, the research has shown that the Information and Communications Technologies opens up new opportunities for transferring medical prescription information securely and faster.

## 5.0 FUTURE WORKS

Future research work would consider the evaluation of the designed system. Also design of an electronic prescription system for specific tropical diseases could also be carried out in later works.

**Note**

This paper is a newer and complete review of: Emuoyibofarhe O.J., Omotosho .A.(2012). Development of a Secure Intelligent E-Prescription System. In proceedings of The International eHealth, Telemedicine and Health ICT Forum For Education, Networking and Business (MedTel 2012) Conference 10th Edition.18-20 April. Luxembourg.

**Author's Brief**

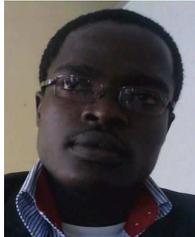

**Omotosho Adebayo** holds a Bachelor of Technology (B.Tech) degree and Master of Technology (M.Tech) degrees in Computer Science both from Ladoke Akintola University of Technology, Ogbomoso, Oyo State, Nigeria. His research areas include e-Health, Telemedicine and Wireless Communication. He can be reached by phone on +2348051867636 and through E-mail bayosite2000@yahoo.com

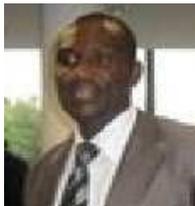

**Emuoyibofarhe .O. Justice** (PhD) is currently a Reader in the Department of Computer Science and Engineering at Ladoke Akintola University of Technology, Ogbomoso, Oyo State, Nigeria. His research areas include Computational Optimisation, Neural Networks, Mobile Computing, Wireless Communication, E-Health and Telemedicine. He can be reached by phone on +2348033850075 and through E-mail eojustice@gmail.com